\documentclass[%
 reprint,
 amsmath,amssymb,
 aps,
]{revtex4-2}
\usepackage{caption}
\usepackage{subcaption}
\usepackage{graphicx}% Include figure files
\usepackage{dcolumn}% Align table columns on decimal point
\usepackage{bm}% bold math
\usepackage[utf8]{inputenc}
\usepackage[T1]{fontenc}
\usepackage{mathptmx}
\usepackage{etoolbox}
\usepackage[english]{babel}

\begin{document}

\preprint{APS/123-QED}

\title{Overcoming the space-charge dilemma in low-energy heavy ion beams \\via a multistage acceleration lens system}% Force line breaks with \\
%\thanks{Contact author: nishiura@nifs.ac.jp}%

\author{M. Nishiura}\email{Contact author: nishiura@nifs.ac.jp}
\affiliation{
National Institute for Fusion Science, 322-6 Oroshi-Cho, Toki 509-5292, Japan\\
Graduate School of Frontier Sciences, The University of Tokyo, Kashiwa, Chiba 277-8561, Japan}

\author{T. Ido}
\affiliation{Department of Advanced Energy Engineering, Kyushu University, Kasuga, Fukuoka 816-8580, Japan}

\author{M. Okamura}
\affiliation{Collider-Accelerator Department, Brookhaven National Laboratory, Upton, New York 11973, USA}
\author{K. Ueda, A. Shimizu, H. Takubo}
\affiliation{National Institute for Fusion Science, 322-6 Oroshi-Cho, Toki 509-5292, Japan}


\date{\today}% It is always \today, today,

\begin{abstract}
Low-energy heavy-ion beams are fundamentally limited by severe space-charge divergence, which constrains the transportable beam current to a few microamperes in conventional electrostatic accelerators. This limitation is particularly critical for high-mass ions, where the generalized perveance increases rapidly because of their low velocity. Here, we demonstrate that this apparent space-charge limit can be overcome by shaping the electrostatic potential configuration of an existing multistage accelerator, thereby transforming the acceleration column itself into a combined acceleration–focusing column. By optimizing the interstage voltage configuration, a strong electrostatic lens effect is superimposed on the accelerating field to counteract the space-charge-driven expansion. We formulate a generalized design framework that quantitatively maps the transport “design window” in terms of beam current, ion mass, and acceleration voltage. For gold ions at 64 keV, this approach enables stable transport of beam currents exceeding 100 µA—more than an order of magnitude higher than the conventional limit. Numerical phase-space analysis shows that this improvement is achieved by prioritizing envelope control over emittance preservation, a trade-off intrinsic to space-charge-dominated regimes. Our results establish a universal and practical guideline for high-current heavy-ion beam transport, relevant to fusion plasma diagnostics, ion implantation, and massive molecular ion applications.

%\begin{description}
%\end{description}
\end{abstract}

\maketitle

%\tableofcontents

\section{\label{sec:intro}Introduction}
The space-charge limit in low-energy heavy ion transport is not merely a technical inconvenience, but a fundamental bottleneck that constrains the scalability of a wide range of accelerator-based applications. Heavy ion beams serve as a cornerstone in a diverse array of applied physics disciplines. Their applications range from Heavy Ion Beam Probe (HIBP) diagnostics for fusion plasma confinement~\cite{Fujisawa_2003, Ido_2010, Nishiura_2024} and intense beam production via ECR ion sources~\cite{Nakagawa_2010} to heavy-ion acceleration in Tandem Van de Graaff systems~\cite{Thieberger_1983}. Beyond fundamental physics, they are critical in industrial Focused Ion Beam (FIB) processing~\cite{Orloff_1993}, material modification via ion implantation~\cite{Ziegler_2010}, and emerging biotechnologies for transporting macromolecules~\cite{Fenn_1989}. In all these applications, maximizing the beam current is imperative to enhance the signal-to-noise ratio or processing throughput.

However, a fundamental physical barrier arises in the low-energy regime (typically tens to hundreds of keV), particularly for high-mass species. **Since the space-charge repulsion scales with the generalized perveance $K \propto (m/q)(I/V^{3/2})$~\cite{Reiser_2008, Lawson_1988}, heavy ions—which possess low velocities due to their large mass ($v \propto m^{-1/2}$)—suffer from severe beam divergence even at moderate current levels.** Consequently, simply increasing the ion source output often fails to improve the transported current, as the Low Energy Beam Transport (LEBT) section becomes the bottleneck where the beam expands and strikes the vacuum chamber walls.

Conventional mitigation strategies primarily rely on magnetic focusing, such as quadrupole magnet arrays. However, the focusing force of a magnetic lens is proportional to the particle velocity ($F \propto v \times B$). Therefore, controlling slow, heavy ions requires prohibitively strong magnetic fields and bulky hardware~\cite{Humphries_1990}, leading to increased system complexity and cost. While electrostatic lenses are an alternative, their design is often constrained by the fixed geometry of existing accelerators.

To address this issue from the perspective of compact device physics, we ask whether this limitation can be overcome solely by actively shaping the potential configuration of an existing multistage accelerator, thereby transforming it into an intelligent optical element. We propose an Active Lensing scheme, in which the multistage acceleration column itself functions as an Active Lens through voltage-only reconfiguration to resolve the space-charge dilemma. Importantly, this concept is realized without introducing any additional external focusing elements to counteract the space-charge repulsion. Here, ``reconfiguration'' refers exclusively to the reassignment of interstage electrode voltages, with the accelerator geometry and hardware remaining completely unchanged. Unlike a conventional Einzel lens, which provides localized focusing in a field-free region, the proposed Active Lens operates by continuously shaping the accelerating field itself, thereby enabling distributed compensation of space-charge-driven divergence throughout the acceleration column.

Furthermore, we present a comprehensive “design window” map that quantifies the physical viability of beam transport. This approach provides a practical and cost-effective guideline for configuring electrostatic acceleration optics, bridging the gap between fundamental beam physics and efficient instrument engineering.

\section{\label{sec:dilemma}The Space-Charge Dilemma in Heavy Ion Transport}

\begin{figure*}[t]
\centering
\includegraphics[width=1\textwidth]{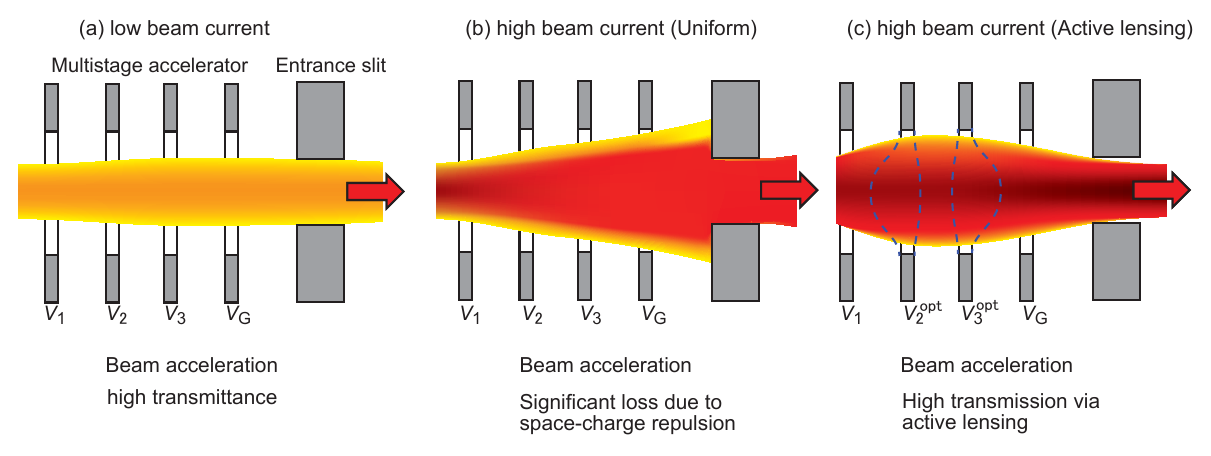}
\caption{\label{fig:space_charge} Schematic illustration of beam envelope evolution within the multistage accelerator column.
(a) Low beam current scenario where space-charge effects are negligible.
(b) High beam current with conventional uniform acceleration ($V_1-V_2 \approx V_2-V_3 \approx V_3-V_G$); significant beam loss occurs at the entrance slit due to space-charge repulsion. (c) High beam current with the proposed active lensing method ($V_{1}-V_{2}^{\mathrm{opt}} \neq V_{2}^{\mathrm{opt}}-V_{3}^{\mathrm{opt}} \neq V_{3}^{\mathrm{opt}}-V_G$). The blue dashed lines indicate the schematic equipotential surfaces formed by the optimized voltage configuration, which exert a focusing force to counteract space-charge expansion and maximize transmission.}
\end{figure*}

\begin{figure}[htbp] 
    \centering
    % --- (a) 上段 ---
    \begin{subfigure}[b]{1.0\linewidth} % カラム幅いっぱいに広げる
        \centering
        \caption{}
        \includegraphics[width=\linewidth]{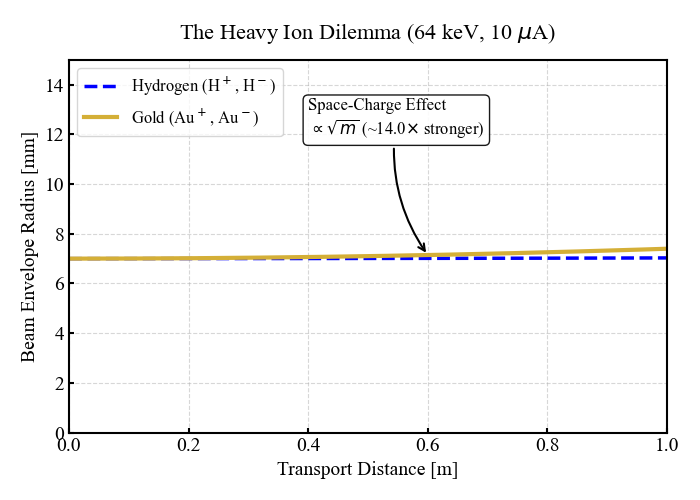}
        \label{fig:envelope_a}
    \end{subfigure}
    
    %\par\bigskip % (a)と(b)の間に適度な余白を入れる

    % --- (b) 下段 ---
    \begin{subfigure}[b]{1.0\linewidth} % カラム幅いっぱいに広げる
        \centering
        \caption{}
        \includegraphics[width=\linewidth]{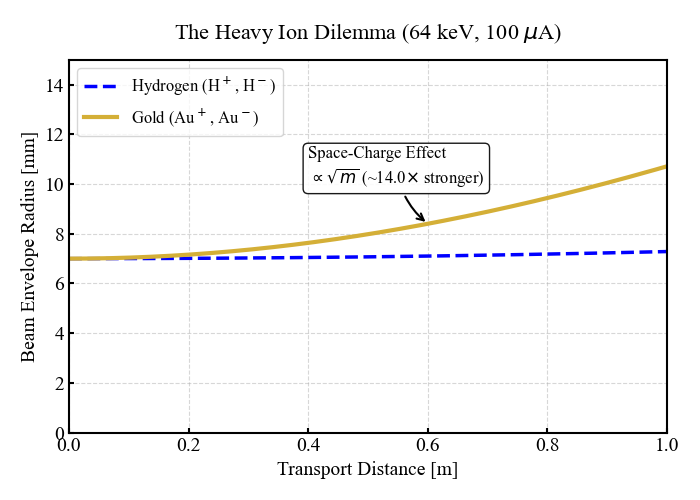}
        \label{fig:envelope_b}
    \end{subfigure}
    
    \caption{\label{fig:envelope} Calculated beam envelope expansion in a field-free drift space for Hydrogen (H$^+$) and Gold (Au$^-$) ions at an energy of 64~keV. (a) Low current case ($10~\mu$A). (b) High current case ($100~\mu$A). While H$^+$ shows negligible expansion, Au$^-$ exhibits significant divergence due to its large mass, highlighting the severity of space-charge effects for heavy ions.}
\end{figure}

In the transport of low-energy, high-intensity heavy ion beams (HIBs), the space-charge effect constitutes a critical hydrodynamic bottleneck. As illustrated in Fig.~\ref{fig:space_charge}(a), trajectory control is trivial in the low-current regime. However, as the beam current increases, the repulsive Coulomb forces dominate, causing significant envelope expansion. Under a conventional uniform acceleration field ($V_1-V_2 \approx V_2-V_3 \approx V_3-V_G$), this expansion results in severe beam loss at the limited acceptance apertures of downstream components, such as the mass-separation slit or the tandem accelerator entrance [Fig.~\ref{fig:space_charge}(b)]. This creates a fundamental dilemma: increasing the ion source current does not translate to a higher transported current, as the geometric acceptance is saturated by the expanding beam halo. Furthermore, relying solely on the pre-acceleration Einzel lens is insufficient because the beam rapidly re-expands during the long drift through the acceleration column.

To quantify this divergence, we consider the beam envelope equation in a drift space. The severity of the expansion is governed by the generalized perveance $K$, a dimensionless parameter representing the ratio of space-charge potential energy to kinetic energy~\cite{Reiser_2008, Humphries_1990}. For a beam of mass $m$, charge $q$, current $I$, and voltage $V_{\mathrm{acc}}$, the radial expansion $\Delta r$ over a drift length $L$ can be approximated as:
\begin{equation}
\Delta r \approx \frac{1}{4\sqrt{2}\pi \epsilon_0} \sqrt{\frac{m}{q}} \frac{I L^2}{V_{\mathrm{acc}}^{3/2} r_0} \propto K \frac{L^2}{r_0},
\label{eq:expansion}
\end{equation}
where $\epsilon_0$ is the vacuum permittivity and $r_0$ is the initial beam radius. Equation~(\ref{eq:expansion}) highlights the intrinsic handicap of heavy ions: since the particle velocity scales as $v \propto m^{-1/2}$ at a fixed energy, the local charge density increases with mass. Consequently, the expansion term scales with $\sqrt{m}$.

The profound impact of this mass dependence is demonstrated in Fig.~\ref{fig:envelope}, which compares the calculated envelopes for Hydrogen and Gold ions over a 1~m drift. While Hydrogen maintains a quasi-laminar flow even at $100~\mu$A ($\Delta r \sim 0.2$~mm), Gold exhibits drastic divergence ($\Delta r \sim 2.8$~mm). This indicates that the space-charge force is approximately 14 times stronger for Gold ($m=197$) than for Hydrogen ($m=1$).

Crucially, Eq.~(\ref{eq:expansion}) also reveals a counter-intuitive ``brightness paradox'': attempting to improve beam quality by reducing the initial radius $r_0$ (e.g., from 10~mm to 7~mm) actually amplifies the divergence ($\Delta r \propto 1/r_0$). Thus, simple geometric collimation is futile. To overcome these physical limitations, we propose the ``Active Lensing'' scheme [Fig.~\ref{fig:space_charge}(c)]. By optimizing the voltages $V_2$ and $V_3$ to their optimal values $V_{2}^{\mathrm{opt}}$ and $V_{3}^{\mathrm{opt}}$, a strong curvature is imparted to the accelerating field, generating an active focusing force that continuously counteracts space-charge repulsion throughout the transport line.

\section{\label{sec:design_window}Design window of low energy beam transport (LEBT) for heavy ion beam}

\begin{figure}
\centering
\includegraphics[width=0.5\textwidth]{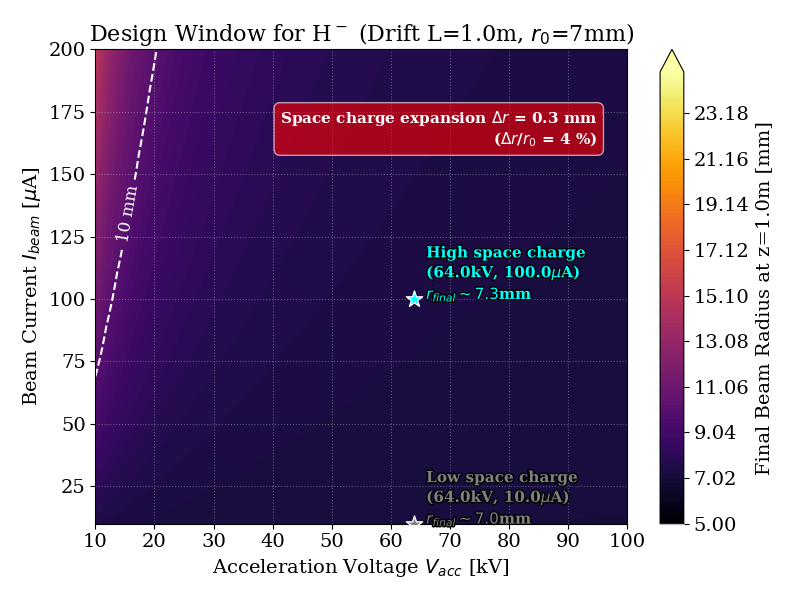}\\
\includegraphics[width=0.5\textwidth]{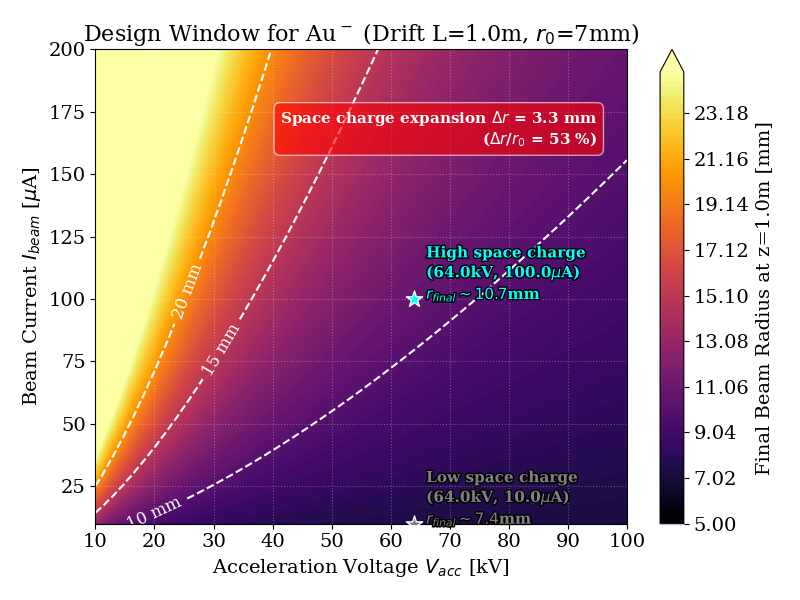}
\caption{\label{fig:fig3_map} Beam design windows accounting for space-charge expansion for (a) hydrogen negative ions ($\mathrm{H}^-$) and (b) gold negative ions ($\mathrm{Au}^-$). The color scale represents the final beam radius after a 1~m drift with an initial radius of $r_0=7$~mm. The star markers highlight the specific cases (10~$\mu$A and 100~$\mu$A) discussed in Fig.~\ref{fig:fig2}.}
\end{figure}

In the acceleration voltage range of 10--100~kV, the behavior of heavy ion beams, such as gold, is strongly governed by space-charge effects. Therefore, to facilitate efficient LEBT beam transport design, we quantitatively mapped the beam radial expansion due to space charge within the parameter space of acceleration voltage and beam current (Fig.~\ref{fig:fig3_map}). Figures~\ref{fig:fig3_map}(a) and (b) show the final beam radii (and expansion rates) for hydrogen and gold negative ions, respectively, calculated based on the aforementioned theoretical formula for a 1~m drift with an initial beam radius of $r_0=7$~mm. The color scale indicates the final beam radius, and the white dashed lines denote isocontours reaching specific beam diameters. 

First, we estimate the beam behavior inside the multistage acceleration tube. We assume the injection of a gold negative ion beam from the source with an energy of 21~keV, 0~rad divergence, and a current of 100~$\mu$A. Within an individual electrode gap ($L=0.05$~m), the beam expansion is approximately $\Delta r \approx 0.05$~mm, rendering the loss in the gap negligible. However, when considering the total length of the multistage acceleration tube ($L \approx 0.5$~m), during which the acceleration voltage gradually increases from 21~kV to 64~kV, the cumulative expansion reaches $\Delta r \approx 2.5$~mm. This suggests that in the high-current regime, space-charge expansion cannot be ignored even within the acceleration tube, necessitating careful electrode design. In contrast, for hydrogen, the beam expansion remains negligible under the same conditions. However, it should be noted that when the beam current reaches the mA order, space-charge forces comparable to those of the gold beam become significant, necessitating similar caution.

Next, we consider the drift region (LEBT) from the acceleration tube exit to the tandem accelerator entrance. For hydrogen (Fig.~3(a)), even in the low-voltage, high-current regime (e.g., 100~$\mu$A) where space-charge forces are most pronounced, the beam radius expansion is limited to $\Delta r/r_0 \approx 4\%$, indicating an extremely broad ``design window.'' In contrast, for gold (Fig.~3(b)), the expansion is drastic due to the mass dependence ($\sqrt{m}$), reaching $\Delta r/r_0 \approx 53\%$ under identical conditions. As indicated by the isocontours (15~mm, 20~mm), the low-voltage regime exhibits rapid beam divergence, expanding the area where the risk of beam loss due to collisions with the beam duct or electrodes becomes unavoidable.

This map provides indispensable guidelines for LEBT design tailored to the specific ion species. Specifically, for heavy ion acceleration, determining the initial values for acceleration gaps, electrode apertures, and downstream component layouts based on this map to ensure the beam is ``physically viable''—prior to conducting computationally intensive numerical simulations—serves as an extremely effective guideline for efficient LEBT design.

\section{\label{sec:optimization}Phase-Space Dynamics and the Trade-off between Emittance and Transmission}
\begin{figure*}[htbp]
    \centering
    \begin{minipage}[t]{0.30\textwidth}
        \vspace{0pt}
        \centering
        \includegraphics[width=\linewidth]{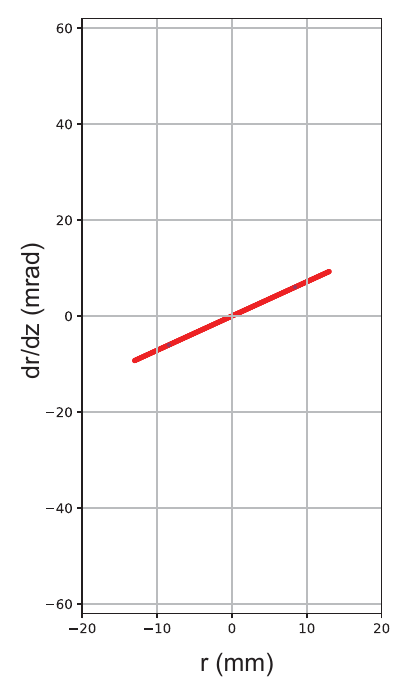}
        \label{fig:phase_space_a}
        \caption*{(a) Initial beam state in front of a multistage accelerator at z = 0 mm.}
    \end{minipage}
    \hfill
    \begin{minipage}[t]{0.30\textwidth}
        \vspace{0pt}
        \centering
        \includegraphics[width=\linewidth]{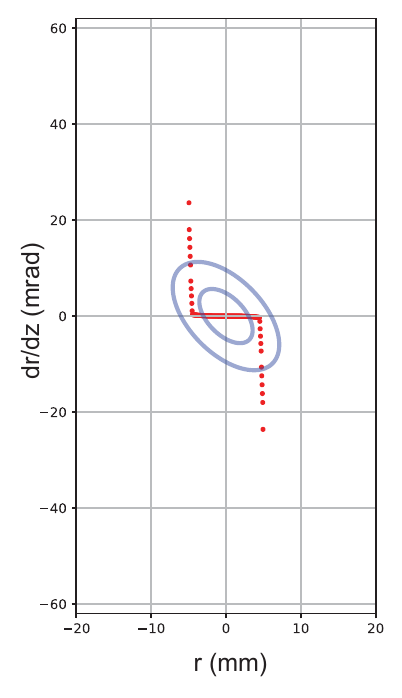}
        \label{fig:phase_space_b}
        \caption*{(b) Uniform acceleration by multistage accelerator with $V_2 = -28$ kV and $V_3 = -14$ kV at $z = $ 1135 mm.}
    \end{minipage} 
    \hfill
    \begin{minipage}[t]{0.30\textwidth}
        \vspace{0pt}
        \centering
        \includegraphics[width=\linewidth]{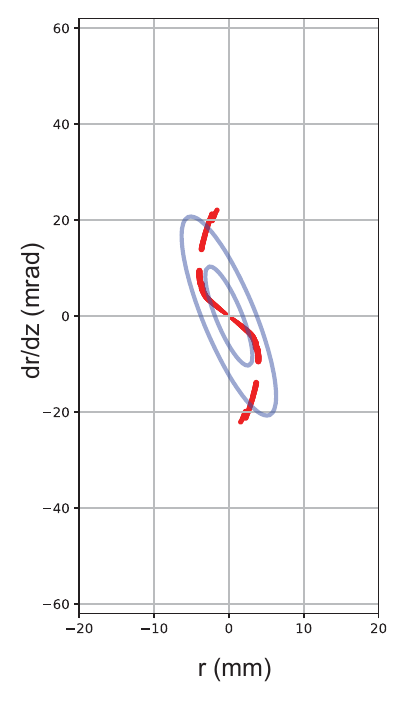}
        \label{fig:phase_space_c}
        \caption*{(c) Active lensing by multistage accelerator with $V_2 = -35$ kV and $V_3 = -3$ kV at $z = $ 1135 mm.}
    \end{minipage} 
    \caption{Phase-space diagrams in a multistage accelerator for Au$^-$ beams with uniform and active lensing method. The inner and outer ellipses indicate $\sigma$ and $2\sigma$ areas.}
    \label{fig:phase_space}
\end{figure*}

Numerical analysis of the beam optics was conducted using the 2D trajectory calculation code IGUN~\cite{Patel:IGUN:2008}. The simulation parameters, including geometry, mesh configuration ($r$ : 250 meshes, $z$ : 1136 meshes) with a grid resolution of 1 mm/mesh and boundary conditions, are consistent with those detailed in our previous work~\cite{Nishiura_2025}. For this study, the initial beam conditions were defined to match experimental values: an initial beam radius of $r_0 = 7$~mm, a gold negative ion beam current of $100~\mu$A, and an initial emittance of $0~\pi$~mm~mrad.

Figure~\ref{fig:phase_space} presents the calculated phase-space distributions (emittance diagrams). Panel (a) shows the initial condition at $z=0$, while (b) and (c) compare the results at the accelerator exit for the uniform acceleration case and the optimized Active Lensing configuration, respectively.

As evident in Fig.~\ref{fig:phase_space}(c), the RMS emittance in the optimized configuration ($\varepsilon_{\mathrm{rms}} \approx 17.6~\pi$~mm~mrad) increases by a factor of approximately two compared to the uniform acceleration case ($\sim 8.5~\pi$~mm~mrad) shown in Fig.~\ref{fig:phase_space}(b). This emittance growth is governed by two physical factors inherent to high-current focusing systems. The first factor is spherical aberration introduced by the strong electrostatic lens. The characteristic ``S-shaped'' distortion observed in Fig.~\ref{fig:phase_space}(c) provides clear evidence that the nonlinear lens fields act disproportionately on the beam periphery, thereby inflating the effective emittance. The second factor is the enhancement of nonlinear space-charge effects. Because Active Lensing strongly compresses the beam envelope, the local charge density increases significantly, intensifying nonlinear Coulomb repulsion and further distorting the phase-space distribution. Such emittance growth is an intrinsic feature of space-charge-dominated beam transport, arising from nonlinear space-charge effects \cite{Lapostolle_1971}. This trade-off is not a drawback but a defining characteristic of optimal transport in space-charge-dominated systems, in which global envelope control governs transport efficiency at the expense of strict emittance preservation \cite{Lapostolle_1987}.

%Such emittance growth is an intrinsic feature of space-charge-dominated beam transport \cite{Lapostolle_1971}. This trade-off is not a drawback but a defining characteristic of optimal transport in space-charge-dominated systems. In space-charge-dominated regimes, envelope control inevitably competes with emittance preservation \cite{Lawson_1988,Lapostolle_1987}.

Although the uniform acceleration case (Fig.~\ref{fig:phase_space}(b)) appears superior at first glance due to its lower emittance, it fails to compensate for space-charge-driven divergence. This design philosophy aligns with the established framework of space-charge-dominated beam transport, in which global envelope control governs transport efficiency \cite{Lapostolle_1987}. In this scenario, the beam envelope expands monotonically, leading to substantial current loss at the transport tube walls. In contrast, the optimized configuration successfully suppresses this divergence by accepting a moderate degree of emittance growth induced by aberrations. This reflects a design philosophy that prioritizes maximizing beam transmission (quantity) over strict emittance preservation (quality)—a rational strategy for the primary objective of delivering high-density current to the target. Consequently, the central current density is enhanced from $25~\mu$A/cm$^{2}$ in the unoptimized case to $75~\mu$A/cm$^{2}$. These simulation results provide robust physical validation for the experimental findings reported in Ref.~\cite{Nishiura_2025}, confirming that the optimization of $V_{2}$ and $V_{3}$ yields a multifold improvement in beam transmission.

\section{\label{sec:application}Transmission efficiency through a tandem accelerator for Heavy Ion Beam}
\begin{figure}
\centering
\includegraphics[width=0.5\textwidth]{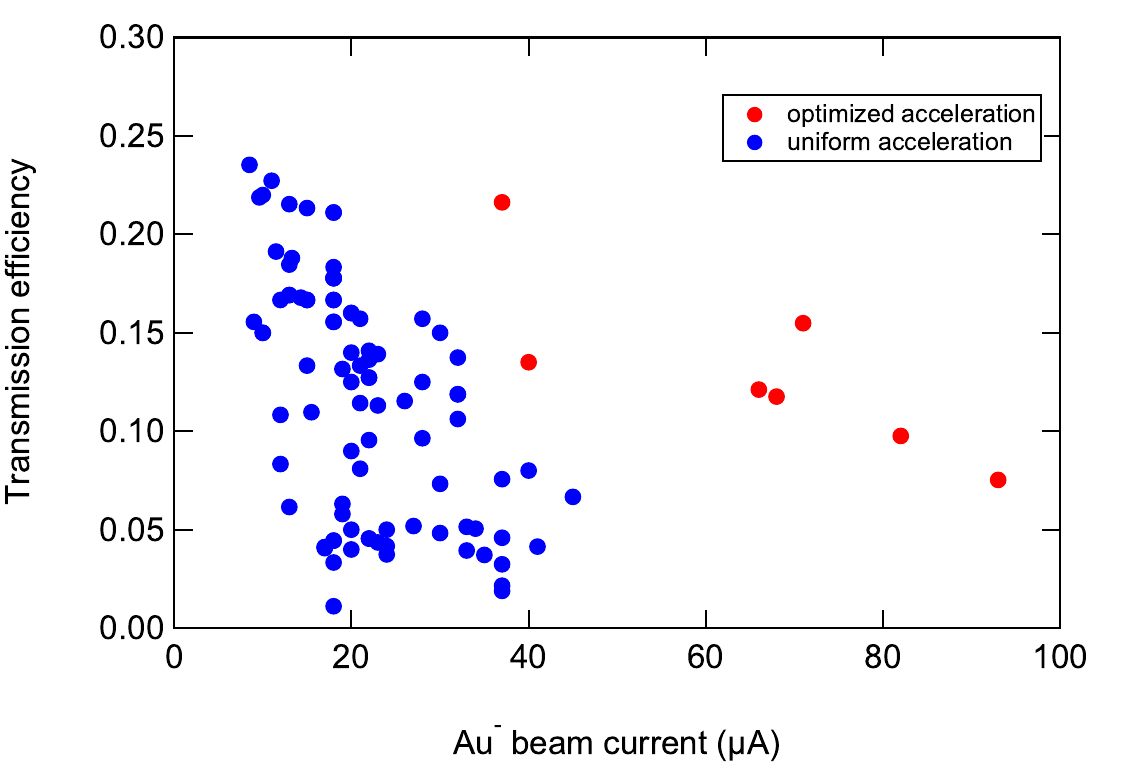}
\caption{\label{fig:fig_transmission_efficiency} Experimental results of the Au beam transmission ($I_{\rm{Au}^+}/I_{\rm{Au}^-}$). Blue circles and red circles denote the uniform-acceleration and active lensing (optimized) -acceleration cases, respectively. The terminal voltage of the tandem accelerator was below 3~MV.}
\end{figure}
Figure~\ref{fig:fig_transmission_efficiency} shows the experimental results of the beam transmission efficiency ($I_{\rm{Au}^+}/I_{\rm{Au}^-}$ input-output ratio) through the tandem accelerator as a function of the injected beam current. We compared the performance of the conventional uniform acceleration (blue circles) with the optimized acceleration using Active Lensing (red circles) applied at the multistage acceleration electrodes prior to injection.

First, the upper limit of the transmission efficiency observed in the low-current regime ($< 10~\mu$A), approximately 24$\%$, is physically determined by the charge-exchange cross-section (from Au$^-$ to Au$^+$) in the gas cell located at the high-voltage terminal of the tandem accelerator.  Based on the theoretical charge-exchange cross-sections reported in Refs.~\cite{Nishiura_2008, Nishiura_2008_NIFS_Rep,BIERSACK198293, Firsov_1959, Shevelko_2004}. We estimated this theoretical transmission limit to be approximately 20$\%$. The general downward trend in transmission efficiency with increasing input current is attributed to space-charge repulsion, which causes the beam to diverge and exceed the accelerator column's geometrical acceptance (physical aperture), leading to beam loss.

It is noteworthy that while the transmission efficiency for the uniform acceleration case (blue circles) degrades rapidly and exhibits scattering beyond $20~\mu$A, the Active Lensing case (red circles) maintains a significantly practical efficiency even in the high-current regime above $40~\mu$A. For instance, at an input current of approximately $90~\mu$A, where the uniform acceleration fails to transport the beam effectively, the optimized condition secures a transmission efficiency of about $8\%$, resulting in a substantial increase in the absolute output current.

The preceding simulation (Fig.~\ref{fig:phase_space}) suggested that the RMS emittance would increase by a factor of two due to lens aberrations associated with Active Lensing, raising concerns that this might hinder beam transport. However, the experimental results demonstrate that even with degraded emittance (beam quality), the effect of Active Lensing—strongly focusing the beam envelope to geometrically match the beam into the acceptance of the tandem accelerator—is dominant in improving transmission. This empirically validates the design philosophy proposed in this study: in heavy ion LEBT systems limited by space charge, prioritizing "suppression of divergence (envelope control)" over "preservation of emittance" is the most effective strategy.

\vspace{\baselineskip}
\section{\label{sec:application}Broader Implications and Future Outlook}

The logical extension of the ``space-charge dilemma'' discussed in this work applies critically to macromolecular ion beams, such as proteins and viral capsids. In emerging fields like Single Particle Imaging (SPI) at X-ray Free Electron Lasers (XFELs), biological samples with masses exceeding $10^5$~amu must be focused into sub-micron interaction regions. Since the space-charge repulsion scales with $\sqrt{m}$ due to the reduced particle velocity at a given energy, the divergent forces in protein beams are over 20 times stronger than in the gold ion beams demonstrated here. Consequently, conventional passive drift transport is rendered completely unfeasible for such massive biomolecules. Our proposed active multistage lens system provides a deterministic solution to transport and focus these fragile, high-mass beams without structural degradation, opening new pathways for native mass spectrometry and ion-soft-landing technologies.

\section{\label{sec:summary}CONCLUSION}
In this study, we comprehensively formulated a heavy-ion beam transport scheme utilizing Active Lensing, generalized across ion species, acceleration voltages, and beam currents, thereby demonstrating its universal utility for LEBT design. While previous work reported significant improvements in transmission efficiency through optimization of multistage acceleration voltages, the significance of this study lies in bridging the gap between existing physical models and practical application.

Specifically, by introducing a ``parameter space map'' that correlates space-charge effects with acceleration parameters, we established a systematic framework for the deductive determination of optimal operating points for arbitrary ion species, providing a seamless connection to detailed design via numerical simulations. Analysis using IGUN provided physical validation for this framework, revealing that the observed performance enhancement arises from a design philosophy intrinsic to space-charge-dominated regimes: strategically prioritizing the suppression of beam divergence (quantity) over strict emittance preservation (quality).

Furthermore, the experimentally achieved transmission reached the theoretical upper limit imposed by the equilibrium yield of the charge-stripping process, demonstrating that the proposed method maximizes performance up to the fundamental physical boundary. Consequently, this work establishes a universal and robust design foundation for high-intensity beam transport systems.

In space-charge-dominated heavy-ion beam transport, the present study demonstrates that optimal performance is achieved by prioritizing global envelope control rather than strict emittance preservation.

\begin{acknowledgments}
This work was supported by the JSPS KAKENHI under Grant Nos. 19KK0073 and 23H01160. We thank the LHD experiment group for their generous experimental support.
\end{acknowledgments}

\section*{Data Availability Statement}
The data that support the findings of this study are available from the corresponding author upon reasonable request.

%\nocite{*}
\bibliography{references}% Produces the bibliography via BibTeX.

\end{document}